\documentclass[preprint,11pt]{elsarticle}

\usepackage{amssymb}
\usepackage{hyperref}
\usepackage{color}
\usepackage{xspace}
\usepackage{comment}
\usepackage{tikz}

\newcommand{\mapp}{\emph{Mapp}\xspace}
\newcommand{\mapps}{\emph{Mapps}\xspace}
\newcommand{\meso}{\textsf{h-MESO}\xspace}
\newcommand{\et}[1]{}  % temporarily silencing comments for arXiv submission

\journal{Current Opinion in Solid State and Materials Science}
\date{}

\biboptions{sort&compress}

\graphicspath{{figures/}{./}}

\begin{document}

\begin{frontmatter}

%% Title, authors and addresses

%% use the tnoteref command within \title for footnotes;
%% use the tnotetext command for theassociated footnote;
%% use the fnref command within \author or \address for footnotes;
%% use the fntext command for theassociated footnote;
%% use the corref command within \author for corresponding author footnotes;
%% use the cortext command for theassociated footnote;
%% use the ead command for the email address,
%% and the form \ead[url] for the home page:
%% \title{Title\tnoteref{label1}}
%% \tnotetext[label1]{}
%% \author{Name\corref{cor1}\fnref{label2}}
%% \ead{email address}
%% \ead[url]{home page}
%% \fntext[label2]{}
%% \cortext[cor1]{}
%% \affiliation{organization={},
%%             addressline={},
%%             city={},
%%             postcode={},
%%             state={},
%%             country={}}
%% \fntext[label3]{}

\title{Integrated Experiment and Simulation Co-Design: A Key Infrastructure for Predictive Mesoscale Materials Modeling}
%% ALT TITLE: \title{Trust Assessment for Mesoscale Modeling}

%% use optional labels to link authors explicitly to addresses:
%% \author[label1,label2]{}
%% \affiliation[label1]{organization={},
%%             addressline={},
%%             city={},
%%             postcode={},
%%             state={},
%%             country={}}
%%
%% \affiliation[label2]{organization={},
%%             addressline={},
%%             city={},
%%             postcode={},
%%             state={},
%%             country={}}

\author[uh]{Shailendra Joshi}

\affiliation[uh]{organization={Department of Mechanical and Aerospace Engineering, University of Houston},%Department and Organization
            addressline={4226 Martin Luther King Boulevard},
            city={Houston},
            postcode={77204},
            state={TX},
            country={USA}}

\author[umich]{Ashley Bucsek}

\affiliation[umich]{organization={Department of Mechanical Engineering, University of Michigan},%Department and Organization
            addressline={2350 Hayward},
            city={Ann Arbor},
            postcode={48109},
            state={MI},
            country={USA}}

\author[psu-mse]{Darren C. Pagan}

\affiliation[psu-mse]{organization={Department of Materials Science and Engineering, Penn State University},%Department and Organization
            addressline={221 Steidle Building},
            city={University Park},
            postcode={16802},
            state={PA},
            country={USA}}

\author[ucsb-me]{Samantha Daly}

\affiliation[ucsb-me]{organization={Department of Mechanical Engineering, University of California, Santa Barbara},%Department and Organization
            addressline={Engineering II},
            city={Santa Barbara},
            postcode={93106},
            state={CA},
            country={USA}}

\author[umn-aem]{Suraj Ravindran}

\affiliation[umn-aem]{organization={Department of Aerospace Engineering and Mechanics, University of Minnesota},%Department and Organization
            addressline={110 Union Street SE},
            city={Minneapolis},
            postcode={55455},
            state={MN},
            country={USA}}

\author[ucla-mse]{Jaime Marian}

\affiliation[ucla-mse]{organization={Department of Materials Science and Engineering, University of California, Los Angeles},%Department and Organization
            addressline={410 Westwood Plaza, 3111 Engineering V},
            city={Los Angeles},
            postcode={90095-1595},
            state={CA},
            country={USA}}

\author[brown-eng]{Miguel A. Bessa}

\affiliation[brown-eng]{organization={School of Engineering, Brown University},%Department and Organization
            addressline={},
            city={Providence},
            postcode={02912},
            state={RI},
            country={USA}}

\author[gatech-me]{Surya R. Kalidindi}

\affiliation[gatech-me]{organization={George Woodruff School of Mechanical Engineering, Georgia Institute of Technology},%Department and Organization
            addressline={801 Ferst Dr},
            city={Atlanta},
            postcode={30332},
            state={GA},
            country={USA}}

\author[uiuc-me]{Nikhil C. Admal}

\affiliation[uiuc-me]{organization={Department of Mechanical Science and Engineering, University of Illinois at Urbana-Champaign},%Department and Organization
            addressline={1206 W Green St},
            city={Urbana},
            postcode={61801},
            state={IL},
            country={USA}}

\author[penn-meam]{Celia Reina}

\affiliation[penn-meam]{organization={Department of Mechanical Engineering and Applied Mechanics, University of Pennsylvania},%Department and Organization
            addressline={220 South 33rd Street, 107 Towne Building},
            city={Philadelphia},
            postcode={19104-6391},
            state={PA},
            country={USA}}

\author[jhu-cee]{Somnath Ghosh}

\affiliation[jhu-cse]{organization={Department of Civil and Systems Engineering, Johns Hopkins University},%Department and Organization
            addressline={3400 North Charles Street},
            city={Baltimore},
            postcode={21218},
            state={MD},
            country={USA}}

%\author[nist]{James A. Warren}
 
%\affiliation[nist]{organization={National Institute of Standards and Technology},%D epartment and Organization
%             addressline={100 Bureau Drive},
%             city={Gaithersburg},
%             postcode={20899},
%             state={MD},
%             country={USA}}

\author[umn-phys]{Jorge Vi\~{n}als}

\affiliation[umn-phys]{organization={School of Physics and Astronomy, University of Minnesota},%Department and Organization
            addressline={116 Church Street SE},
            city={Minneapolis},
            postcode={55455},
            state={MN},
            country={USA}}

\author[umn-aem]{Ellad B. Tadmor\corref{cor1}}
\cortext[cor1]{Corresponding author, \url{tadmor@umn.edu}}

% AUTHORS
% OK  Nikhil Chandra Admal <admal@illinois.edu>
% OK  Ashley Bucsek <abucsek@umich.edu>,
% OK  Surya Kalidindi <surya.kalidindi@me.gatech.edu>,
% OK  Darren Pagan <dcp5303@psu.edu>, 
% OK  Suraj Ravindran <sravi@umn.edu>,
% OK  Sam Daly <samdaly@engineering.ucsb.edu>,
% OK  Celia Reina <creina@seas.upenn.edu>,
% OK* James Warren <james.warren@nist.gov>, 
% OK  Jaime Marian <jmarian@g.ucla.edu>,
% no  Krishna Garikipati <garikipa@usc.edu>, (just acknowledge)
% OK  Miguel Bessa <miguel_bessa@brown.edu>
% OK  Somnath Ghosh <sghosh20@jhu.edu>
%     Suraj Ravindran
%
% * Needs to go through NIST approval. OK to put on arXiv without Jim, and 
%   add him in a new version once approved

\begin{abstract}
The design of structural and functional materials for specialized applications is experiencing significant growth fueled by rapid advancements in materials synthesis, characterization, and manufacturing, as well as by sophisticated computational materials modeling frameworks that span a wide spectrum of length and time scales in the \textit{mesoscale} between atomistic and homogenized continuum approaches. This is leading towards a systems-based design methodology that will replace traditional empirical approaches, embracing the principles of the \textit{Materials Genome Initiative}. However, there are several gaps in this framework as it relates to advanced structural materials development:
(1) limited availability and access to high-fidelity experimental and computational datasets,
(2) lack of co-design of experiments and simulation aimed at computational model validation,
(3) lack of on-demand access to verified and validated codes for simulation and for experimental analyses, and
(4) limited opportunities for workforce training and educational outreach. These shortcomings stifle major innovations in structural materials design. This paper describes plans for a community-driven research initiative that addresses current gaps based on best-practice recommendations of leaders in mesoscale modeling, experimentation, and cyberinfrastructure obtained at an NSF-sponsored workshop dedicated to this topic and subsequent discussions. The proposal is to create a hub for \textit{Mesoscale Experimentation and Simulation co-Operation} (\meso)---that will (I) provide curation and sharing of models, data, and codes, (II) foster co-design of experiments for model validation with systematic uncertainty quantification, and (III) provide a platform for education and workforce development. \meso will engage experimental and computational experts in mesoscale mechanics and plasticity, along with mathematicians and computer scientists with expertise in algorithms, data science, machine learning, and large-scale cyberinfrastructure initiatives.
\end{abstract}

%%Graphical abstract
%\begin{graphicalabstract}
%\includegraphics{grabs}
%\end{graphicalabstract}

%%Research highlights
%\begin{highlights}
%\item Research highlight 1
%\item Research highlight 2
%\end{highlights}

\begin{comment}
\begin{keyword}
%% keywords here, in the form: keyword \sep keyword
keyword one \sep keyword two
%% PACS codes here, in the form: \PACS code \sep code
\PACS 0000 \sep 1111
%% MSC codes here, in the form: \MSC code \sep code
%% or \MSC[2008] code \sep code (2000 is the default)
\MSC 0000 \sep 1111
\end{keyword}
\end{comment}

\end{frontmatter}

%% \linenumbers

%% main text
\section{Introduction}
\label{sec:intro}
\textit{In silico} design of structural materials has been fueled by four key drivers: (1) the need to develop a systems-based approach  \citep{MPC09} for rapidly designing materials with tailored and reproducible mechanical responses as envisioned in the Materials Genome Initiative (MGI) \citep{MGI11}, (2) the advent of superior material synthesis and processing platforms such as additive manufacturing, (3) the development of advanced experimental techniques for three-dimensional (3D) material characterization along with predictive physics-based models for microstructure-sensitive modeling and simulation \citep{McD18}, and (4) the rise of cyberinfrastructures that promote collaborative work by sharing software and data, and that enable systematic assessment of models through the application of data analytics and advanced statistical and learning-based approaches \citep{TEP13}.The 2021 MGI strategic plan \citep{MGI21} focuses on harnessing these drivers toward the goal of a unified infrastructure for materials innovation.

For structural materials design integrated with advanced manufacturing, such a program would naturally require the development of integrative experimental and computational protocols for process--microstructure--property linkages that enable accurate and efficient calculations of mechanical response and failure envelopes accounting for microstructural evolution \citep{Kal15}. A prerequisite to successfully realizing this vision is a robust experimental and computational mechanics of materials infrastructure at the \textit{mesoscale}\footnote{We define ``mesoscale'' as length and time scales that bridge atomistics and fully homogenized continuum mechanics.} that resolves the key microstructural constituents, provides a systematic framework for uncertainty quantification (UQ), and defines standardized protocols for verification and validation (V\&V). An insightful study sponsored by the U.S.\ National Science Foundation (NSF) on computational modeling in the mechanics of materials \citep{TMS19} identifies V\&V and UQ as research priorities, and points out that the current research infrastructure in this domain faces significant gaps that hinder the vision of accelerated material discovery through integrated modeling and experimentation. These include:
\begin{itemize} \setlength\itemsep{-0.5ex}
\item[(1)] Limited availability and access to high-fidelity experimental and computational datasets that span length and time scales and which are necessary in the development of new computational models.
\item[(2)] Lack of co-design of experiments and simulation, i.e., designing experiments that can be used to parameterize and validate computational models.
\item[(3)] Lack of on-demand access to verified and validated codes for simulation and for experimental analyses, which includes the development and sharing of rigorous V\&V and UQ protocols.
\item[(4)] Limited opportunities for workforce training and educational outreach providing the necessary background in V\&V, UQ, data science, and software carpentry skills.
\end{itemize}
A followup study \citep{TMS20}, also sponsored by NSF, envisions mechanisms by which these and related barriers can be tackled and provides exemplars across several problems in mechanics and materials.

\begin{figure}[t!]
    \centering
    \includegraphics[width=0.9\textwidth]{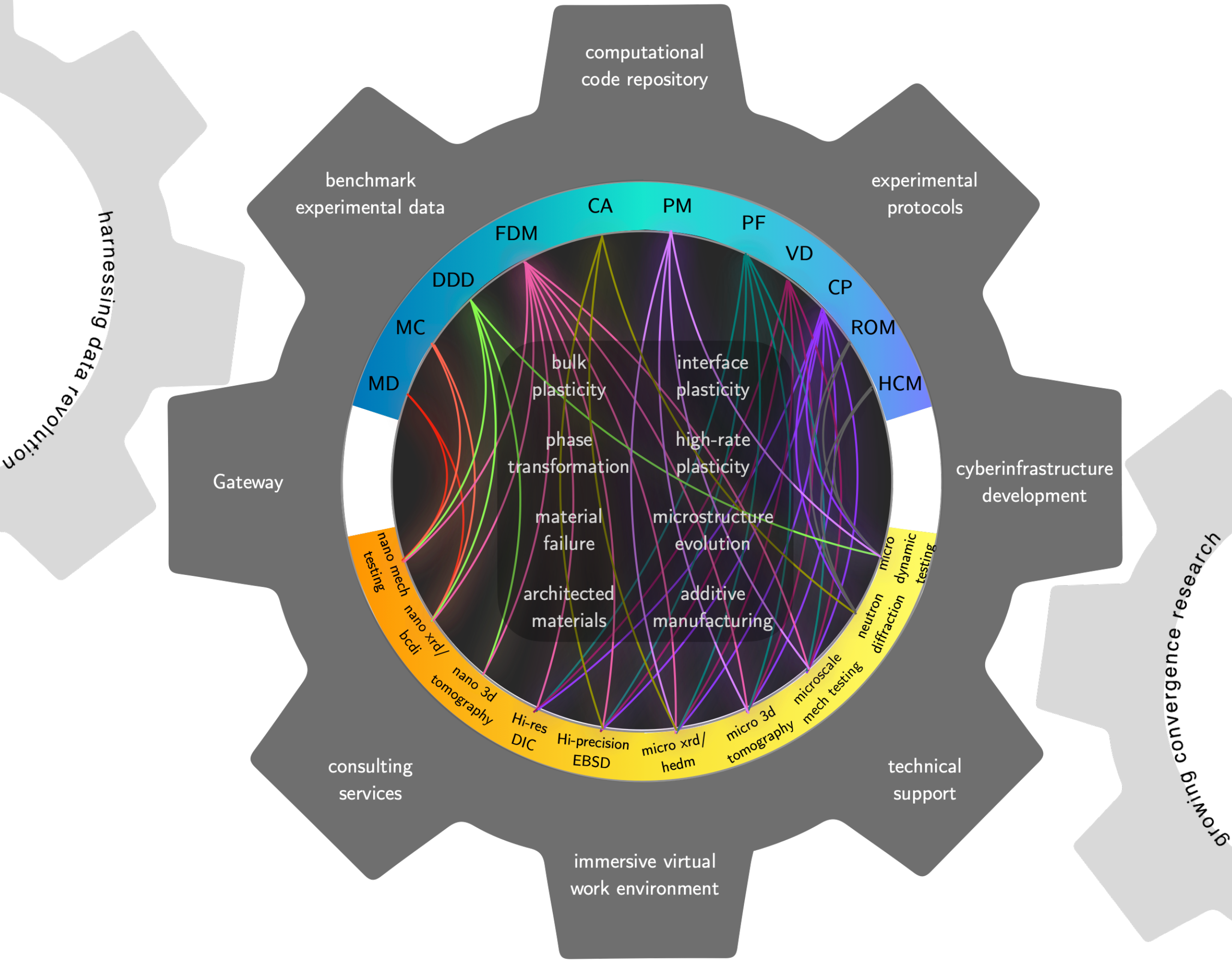}
    \caption{The \meso infrastructure integrates computational and experimental domains in the mechanical behavior of crystalline materials by engaging researchers across different disciplines, developing a scalable research infrastructure with an immersive virtual working environment, creating and curating verified and validated simulation codes (\mapps), recommending standardized experimental and computational protocols, offering technical support and consulting services, and offering a streamlined gateway for managing and sharing workflows, histories, and metadata. The acronyms in the figure are different \mapps: MD (molecular dynamics), MC (Monte Carlo), DDD (Discrete Dislocation Dynamics), FDM (Field Dislocation Mechanics), CA (Cellular Automata), PM (Particle-based Methods), PF (Phase-field), VD (Vertex Dynamics), CP (Crystal Plasticity), ROM (Reduced Order Models), HCM (Homogenization-based Continuum Plasticity). 
%\meso connects to two of the NSF 10 Big Ideas.
    }
    \label{fig:KnoMME}
\end{figure}

This paper represents a community view on a way forward based on an NSF funded workshop held in 2023 and followed by continued discussions in different forums. This effort led to the conclusion that the challenges described above can be addressed, gaps bridged, and the field of mesoscale materials modeling advanced, through the creation of a community-driven interdisciplinary research hub for \textit{Mesoscale Experimentation and Simulation co-Operation} (\meso)\footnote{This name reflects the intended cooperation between experimentalists and theorists performing simulation, the emphasis on co-design of experiments and simulations, and the fact that this would be a dynamic ongoing operation.}, Fig.~\ref{fig:KnoMME}. \meso would comprise a scalable research infrastructure engaging domain experts in experimental and computational mesoscale mechanics, as well as experts from other fields that inform such an effort including engineers, mathematicians, and computer scientists with expertise in UQ and data-driven science.
%The envisioned effort aligns with the NSF's 10 Big Ideas \citep{nsftenbig}: it is a \textit{Mid-scale RI} initiative with strong elements of \textit{Growing Convergence Research} and \textit{Harnessing the Data Revolution}.

\section{Background}
\label{sec:background}

\subsection{Mesoscale Modeling and Simulation}
\label{sec:meso:modeling}
Understanding, predicting, and controlling the mechanical behavior of structural materials is at the core of designing strong and damage-tolerant structural components–from microelectronic devices to aircraft. Plastic deformation and failure are arguably the most significant processes guiding such application oriented materials design and development. In metals, plasticity is rich with phenomena (e.g., dislocation evolution, deformation twinning, and stress-induced phase transformations) that occur and interact on multiple length and time scales; the choice of a modeling framework depends on the extent to which these phenomena need to be resolved in a given problem. At the finest scales, atomistic modeling using molecular dynamics (MD) probes the fundamental unit processes of plasticity and their interactions \citep{ZepedaRuiz2017, TM11}. At the other extreme, the plasticity of macroscale components is modeled using homogenization-based continuum plasticity (HCM) theories that are agnostic to finer microstructural details (e.g.,~J$_2$ flow theory) \citep{GFA10}.

Modeling frameworks spanning this broad range of scales are based on mesoscale deformation theories with varying levels of discreteness, which include, discrete dislocation dynamics (DDD), field dislocation mechanics (FDM), phase-field (PF), and crystal plasticity (CP) among others \citep{tadmor:phillips:2000,BAK98,NED00,ACH04,BH16,PMC14}. More recently, mesoscale plasticity modeling approaches based on machine learning/data science are rapidly gaining interest in the community \citep{BBL17, MBC19, MSK21}. These mesoscale frameworks are critical to understanding structure-property linkages for a wide spectrum of materials processing (e.g., severe plastic deformation, sheet metal forming, casting, extreme loading conditions) and advanced manufacturing (e.g., additive manufacturing, 3D printing, and micro- and meso-machining) platforms.

We refer to the computational implementations of mesoscale approaches as \mapps (see Fig.~\ref{fig:KnoMME}). Each \mapp is itself a class of methods that in turn bridge continuum scale descriptions of plasticity with the kinetics of deformation mechanisms at finer scales, thereby providing an understanding of plasticity over a range of length scales from sub-micron to millimeter scales \citep{DM97, Daw00, NG01, CDF16, IBJ20} and time scales from dynamic impact to creep \citep{REH10, McD10}. Moreover, a wide gamut of microstructure-sensitive phenomena can be addressed, such as mechanical instabilities \citep{ASA83}, fatigue \citep{MD10, HE16}, damage \citep{rmo11, SJB19, JJ22, IJB22}, irradiation behavior \citep{BAM13}, and recrystallization and grain growth \citep{APM18}.

\subsection{Mesoscale Experimentation}
\label{sec:meso:experimentation}
\et{Need to expand a bit to match the size of the other background sections; 1--2 more paragraphs}
In concert with computational progress, recent advances in experimental techniques for direct measurement and observation of mesoscopic plastic phenomena provide unprecedented opportunities for validating computational models and thereby improving process-microstructure-property linkages \citep{CT18}. These groundbreaking techniques include nano 3D tomography \citep{nano3dtom_1,nano3dtom_2}, 3D Bragg coherent diffraction imaging (BCDI) \citep{Sut17,bcdi_1}, high-resolution digital image correlation (Hi-res DIC) \citep{hiresdic_1,hiresdic_2}, high-precision electron backscattering diffraction (EBSD) \citep{WNF11,highprecebsd_1,highprecebsd_2}, dark field X-ray microscopy \citep{SKL15}, high-energy X-ray diffraction microscopy (HEDM) \citep{SMO19, MVB09, GLS20,hedm_1,hedm_2}, ultrafast in-situ scanning and/or transmission electron microscopy \citep{LGM20}, micro 3D tomography \citep{microtom_1,microtom_2}, nano and micromechanical testing platforms \citep{nanomech_1,nanomech_2, microtest_1,microtest_2}, and dynamic testing platforms \citep{dyntest_1,dyntest_2}. These modern experimental techniques are characterized by extremely large datasets that require specialized data science knowledge to effectively process, analyze and mine for knowledge.

\subsection{Materials Cyberinfrastructure}
\label{sec:background:ci}

\et{This section should be shortened.}
Tighter integration between experimentation and data management on the one hand, and modeling and simulation on the other, is expected to have a significant impact in materials science research. An optimal path to achieve this goal is to develop and promote an advanced cyberinfrastructure \cite{TEP13} to facilitate frictionless information flow during the repeating cycle of theory and experimentation. Common steps in the cycle involve data acquisition, storage, analysis and curation, concurrently with computational model development and validation. A properly designed cyberinfrastructure, for example, in the form of a Science Gateway \cite{Lawrence2015-mq,Wilkins-Diehr2018-rx,Gesing2023-ho}, would allow optimal use of data sets by the broad community, and the ability to combine them more effectively with theory and simulation.

Raw data sets from modern facilities are often costly in terms of resource requirements, but uniquely rich in content, which is, unfortunately, not always fully utilized. Data sets are also large in size, and therefore difficult to transport and store. In essence, processed output data sets track 3D structural responses to a variety of imposed conditions. They are then mined to find specific classes of events (e.g., twinning, slip, coarsening, phase transformations), so as to determine the conditions under which they occurred, and to elucidate the physical mechanisms behind them. But perhaps one of their most potentially valuable use is to help in the development and validation of computational models of these specific events, and of the overall evolution of structure. Conventional work along the theory-experimentation integration path \cite{Lind2014,Pokharel2015,Pokharel2014}, has generally been confined to one-off comparisons motivated by a specific research question, rather than in depth, repeated interactions between model and experimental data. More recent work \cite{Pagan2017,Turner2017,Chatterjee2016,Chatterjee2017,Tari2017} has successfully addressed, over limited ranges of processing conditions, observed responses through tuning of model parameters or even the models themselves in order to describe specific material phenomena. We suggest that the time is ripe for the development of a cyberinfrastructure framework that would make such comparisons more straightforward, indeed routine, so that many researchers from different backgrounds can effectively access both theory and data. 

Current challenges, described in more detail in Section~\ref{sec:meso:challenges}, include the selection of key phenomena and ancillary benchmark datasets that would be used in the design of an initial implementation of the \meso infrastructure, the definition, development, and integration of standardized tools for the reduction and analysis of large data sets acquired at university laboratories or national facilities, the ingestion of these data sets into common formats appropriately typed and containing provenance and metadata information, workflow managers linking the data to imaging, and modeling codes, integrated connectivity to large scale storage and computational resources in which to run the models, and the definition of the architectural details of the gateway including specific information about deliverables and metrics of adoption and success. At present, no science gateway exists for the large segment of the materials science community whose research revolves around structural properties of advanced materials as revealed by high resolution and in-situ diagnostics (high energy X-ray diffraction and tomography, electron microscopy and backscatter diffraction, digital image correlation), and related theory and simulation.

The following steps are envisioned in the functioning of the gateway. First, data from heterogeneous sources at different locations needs to be ingested. For very large data sets, it is probably preferable that they be left in the storage platforms of the facilities in which they were created. There exist mechanisms  to make this data accessible within a workflow manager for downstream analysis. Second, the data is ingested into the gateway implementation, typically as an automated process using a web API. During this second stage, the data is typed, undergoes format validation, and metadata is attached to facilitate workflows and to ensure that efficient high-performance computing (HPC) resources are used during subsequent analyses. The data becomes immediately available to gateway users from a shared data library; access permissions can also be set to restrict use to specific users. Formats for output data sets to facilitate interactions with and comparisons to computational models are part of the design of the gateway, and are enforced in its implementation. More details on Science Gateway infrastructure and management is provided in \ref{sec:sgmanagement}.

\subsection{Co-Design Challenges}
\label{sec:meso:challenges}
While the literature on \mapp precursors is rich and dates back many decades, efforts to advance the field by leveraging enhanced computational capabilities and improved physical understanding through advanced experimentation have been fragmented, lacking a unifying framework within which to assess their validity and transferability. Further, the successful application of \mapps in the microstructure-sensitive design of materials hinges on how faithfully they represent the physics of plasticity across multiple length and time scales \citep{TPO00}. Several published works point to the challenge of validating full-field predictions of micro-mechanical response \citep[e.g.,][]{ PLK14}. {\it A more robust and systematic co-design effort integrating modeling and experiments is needed to establish reliable connections across the scales to accomplish true scale bridging and/or parameter-less \mapp frameworks} \citep{MZH16, TLP18, AA20a, AZA20, AA20b, CPZ20, CRL21}. This may prompt discussion on the quality and applicability of physical models of deformation, model uncertainty, transferability, and validation \citep{ritchie20}. A comprehensive and accessible infrastructure combining mesoscale plasticity modeling, material parameterization, diagnostic comparison (V\&V), UQ, and data mining does not exist.

 Such fragmentation and lack of coordination present challenges for the materials research community:
 
 \paragraph{Modeling challenges:}
\begin{itemize}  \setlength\itemsep{-0.5ex}
    \item \mapps  are generally not readily available for on-demand use in community codes thereby limiting their usage. Further, the lack of centralized curation of \mapps subject to provenance control makes reproducibility of others' work difficult or impossible.
    \item Typical research codes are designed to perform well for particular classes of research problems, with limited transferability to other problems. The range of applicability of codes is often not widely known or reported.
    \item Research codes are often poorly documented, and are not normally developed using state-of-the-art software engineering practices (DevOps), including version control, implementation verification, continuous integration and deployment, scalability, and adaptability (e.g., ease of incorporating new features).
\end{itemize}

\paragraph{Experimental challenges:}
\begin{itemize}\setlength\itemsep{-0.5ex}
    \item Datasets generated in advanced mesoscale experiments are typically extremely large and are difficult to share with the research community. As a result, many datasets remain unavailable.
    \item There is insufficient emphasis on the design of standardized experiments, and the repeated performance of such experiments by different groups and at different facilities to ensure reproducibility and assess uncertainty.
    \item Analysis of mesoscale experiments require specialized codes and knowledge to process large datasets, apply statistical and informatics tools, perform model reduction, and extract features that can be compared with models. There is currently limited sharing of such tools between experimental groups and facilities.
\end{itemize}

\paragraph{Integration challenges:}
\begin{itemize}\setlength\itemsep{-0.5ex}
    \item Published datasets tend to be only a small fraction of the vast amount of information collected from computations and experiments. In particular, negative results, as well as soft knowledge and gained insights may not be shared. As such, much of the data remains grossly underutilized and may not be easily accessible to the community for further analysis.
    \item Systematic calibration and validation of \mapps against experiments at appropriate length-scales and time-scales is hindered by a disconnect between the data required for model calibration and assessment, and the results obtained from experiments that were not designed with those needs in mind.
    \item There is a shortage of workforce with strong computational expertise and domain knowledge integrating advance computing, data science, and machine learning with mesoscale plasticity and failure modeling and experiments.
    \item The lack of easy access to codes and experimental data is a problem for the community as a whole but is particularly challenging for new researchers attempting to enter the field.
\end{itemize}

These issues point to the need for a universal research infrastructure---\meso---to serve as an online open resource for the curation of \mapps, standardized experimental protocols and associated computational tests designed to validate them, and associated experimental data and software tools. While efforts aimed at creating a modular and an open research infrastructure for materials innovation exist, they are primarily designed for atomistic modeling, e.g., the OpenKIM effort \citep{tes11}. The few efforts that exist in the realm of mesoscale modeling are somewhat limited in scope \citep{REK12, PRISMS}.

\section{\meso in Broad Strokes}
\label{sec:aims}
The \meso research infrastructure (Fig.~\ref{fig:KnoMME}) would be functionally organized, and presented to the broader community, as an online interdisciplinary research center with associated computing and data storage resources.
%The center comprises
The center will be built around
physics-based computational applications at the mesoscale, referred to as ``Mesoscale Applications'' or \mapps (shown in the upper blue ring in Fig.~\ref{fig:KnoMME}) that are connected with experimental methods that can be used to validate them (shown in the yellow ring in Fig.~\ref{fig:KnoMME}). Both the \mapps and experimental methods are ordered from short length and time scales on the left to long scales on the right. Each \mapp and experimental method are themselves a class of methods (e.g., the CP \mapp consists of many varieties of crystal plasticity models). At the start,  \meso will focus on a small number of \mapps, each with its own combination of benchmark experimentation, theory, and associated computation. Based on community interest, additional \mapps will be added over time.

The ensemble of \mapps and experiments will be supported by a common infrastructure.  The same structural connection between experimental and theoretical activities will be replicated across \mapps and supported by a common workflow manager and web-facing gateway.
%The ensemble of \mapps and experiments will be connected and supported through a common infrastructure comprised of a workflow manager and web-facing gateway.
This will include the definition of standard \mapp and experiment outputs similar to ``KIM Property Definitions'' \citep{kimprop} employed by the OpenKIM Processing Pipeline \citep{karls:bierbaum:2020, karls:clark:2022}, and the development of translation tools such as those used by the Galaxy workflow manager \citep{Galaxy_complexdata,Galaxy_interoperability,Galaxy_unified,galaxy,miller2022}. This will facilitate translation of research across \mapps, and seamless navigation by outside parties who will become users of the center. It will also make it straightforward to add new computational and experimental methods to \meso.

The common infrastructure of the center will provide and support computational resources in which to develop, deploy, and use the computational codes associated with \mapps, experiments and pre- and post-processing data analysis.
%developed, maintained, and curated within each \mapp.
Managed storage will contain data from the benchmark experiments conducted as part of the project, as well as derived data, histories, and workflows from downstream analysis and computation. \meso will leverage recent developments in virtual reality (VR) \citep{schaefer:reis:2022} to create an immersive environment for collaboration and education. VR equipment will be provided to external participants through a merit-based application process, also aimed at increasing participation of underrepresented groups.

Each year, \meso researchers will organize in workgroups focused on specific physical themes (see examples in the center of the circle in Fig.~\ref{fig:KnoMME}) in which theorists and experimentalists will work together to design simulations and experimental protocols to validate them. These efforts will culminate in validation challenges, as advocated in the NSF V\&V reports \citep{TMS19,TMS20}, integrated within a summer program in which fully-funded external participants will be attempt to reproduce experimental results using their \mapps and receive training and educational content.\footnote{The summer validation challenges will be similar to the successful Sandia Fracture Challenge \citep{KJM19} and the NIST Nanoindentation Round Robin \citep{RKB07} events in which modelers were challenged to predict results of specially-designed experiments.}

The center will also feature a complement of advanced disciplinary scientists tied to the various \mapps who will provide both internal and external consulting services and support. A small team of information technologies (IT) professionals and computer scientists will be responsible for managing the infrastructure, supporting the development of machine learning and data science tools associated with \mapps and experimental data processing, and coordinating VR activities and developing VR-based content.

Finally, the center's scientific personnel will provide support not only to members of the project, but to the wider community. Dedicated, in-depth support, will be possible through cost sharing arrangements with external users of the facility. This will allow for quick dissemination of the knowledge developed within \meso, as well as efficient training of new users in the utilization of resources provided by the infrastructure.

\section{Community Best-Practice Recommendations for \meso}
\label{sec:community}

%\sj{I have added the workshop agenda to the appendix. It is creating some arrangement issues, but I will fix them.}

To identify community needs and define best practices for the design of the envisioned \meso infrastructure (see Section~\ref{sec:aims}), a workshop funded by the NSF was held January 12--13, 2023 at the University of Houston in collaboration with the University of Minnesota. The workshop included invited participants identified as leaders in the mesoscale modeling, experimentation, and cyberinfrastructures (see \ref{app:participants} for a list of participants). The workshop included thematic talks from recognized experts, followed by panel discussions focused on three themes:
\begin{enumerate}
\item Nature of mesoscale models
\item Experimental validation
\item Cyberinfrastructure vision and design
\end{enumerate}
Participants were divided into workgroups organized around the themes. In each theme, participants were provided with a set of questions prepared by the conference organizers (Ellad Tadmor and Shailendra Joshi) and asked to provide best-practice recommendations. The questions, and committee responses following deliberation are given below.

The workshop keynote address was given by Prof.~Graham Candler from the University of Minnesota who was involved in a successful effort to validate computational fluid dynamics (CFD) codes for hypersonic applications against experimental wind tunnel experiments \citep{Knight2012-mb, Ray2020-pb}. This talk and subsequent discussion provided the participants with insights from the experiences of the CFD community that helped in their deliberations.

\subsection{Nature of Mesoscale Models}
\label{sec:theme1:meso}

\subsubsection{What is the role of mesoscale modeling and \mapps in scientific and technological progress within the realm of materials design?}

\medskip
The mesoscale captures length and time scales over a broad range between atomistic and homogenized continuum treatments. It is widely recognized in the materials community that a deep understanding of materials behavior at the mesoscale can have a transformative impact on materials discovery, materials development acceleration, and deployment into new technologies and applications.

\begin{table}
\begin{tabular}{lcc}
Method & General- & Phenom- \\
       & purpose & focused \\
\hline
CA (Cellular Automata) & \checkmark & \\
CP (Crystal Plasticity) & & \checkmark \\
DDD (Discrete Dislocation Dynamics) & & \checkmark \\
FDM (Field Dislocation Mechanics) & & \checkmark \\
HCM (Homogenization-based Continuum Plasticity) & & \checkmark \\
MC (Monte Carlo) & \checkmark & \\
MD (molecular dynamics)  & \checkmark & \\
PF (Phase-field) & \checkmark & \\
PM (Particle-based Methods) & \checkmark & \\
ROM (Reduced Order Models) & \checkmark & \\
VD (Vertex Dynamics) & \checkmark & \\
\hline
\end{tabular}
\caption{Identification of \mapps as general-purpose or phenomenon-focused methods.}
\label{tab:mapps}
\end{table}

\mapps can be categorized as general-purpose methods with wide application fields or as phenomenon-focused methods whose development is specifically driven by an application as shown in Table~\ref{tab:mapps}.

%\item Or perhaps \mapps should organized around well defined, clean, physical problems (test %problems).
%\begin{itemize}
%\item Single crystal: binary dislocation junction formation; nanopillar deformation, nanoindentation tests
%\item Bi-crystal: subsets of 5D GB energy surface at 0K, dislocation/GB interactions
%\item Poly-crystal: Hall-Petch, SRX, DRX, GB mobilities, triple junction behavior
%\item Classification in terms or what solvers do they use (implementation, flexibility)
%\item Taxonomy for interrelations: Test problems $\rightarrow$ method $\rightarrow$ implementation $\rightarrow$ data (ML) $\leftarrow$ experimental data
%\end{itemize}
%\end{itemize}

\subsubsection{What capabilities do these models bring that are absent in atomistic or continuum approaches?}

\medskip
There is a broad range of physical processes operating on scales that are inaccessible to atomistic methods on the one hand, while also being too complex for ``homogenized'' continuum approaches based on mean-field approximations. Material behavior at this \textit{mesoscale} are crucial for many emerging applications and technologies.

\subsubsection{Who are the stakeholders in mesoscale modeling in academia, industry, and government and what are their needs?}

\medskip
There are a variety of different stakeholders with different needs. In academia, a key constinuency are researchers working towards specific code validation, or those interested in finding a specific computational tool that performs well for a given problem of interest. Experimentalists benefit from validated \mapps to help interpret experiments, and can use a \meso infrastructure to curate, and release large data sets. The latter is critical, as scientifically-valuable data (of any type or source) should be of the highest quality, openly available, and reproducible. This should be assured through discussion with government agencies (through funding requirements) and agreements with publishers and editorial policymakers to ensure wide access to data, which do \textit{not} rely on personal altruism.

Government buy-in for efforts like \meso will depend on each agency's mission statement (NIST, DOE, NSF, etc). However, broad community consensus will always be needed for government support. Therefore meetings like the workshop that led to these recommendations are a necessary first step.

Industry stakeholders are likely to be interested in suites of tools to aid them in materials design that cover the relevant scales of interest. To be widely adopted such \mapps should be easy to use, well-documented and integrated as promoted by current Integrated Computational Materials Engineering (ICME) approaches \citep{Horstemeyer2018-lh} (see Section~\ref{sec:icme}).

%\subsubsection{What are the inputs and outputs of different \mapps and which phenomena do they model?}
%\begin{itemize}
%\item Define a `RVE model' that captures the mesoscopic properties of a component and allow complex components to be studied with relative ease.
%
%\item Employ formatting protocols that are either in existence and in wide use, or develop new ones with user convenience in mind.
%
%\item Optimize the utility of input definitions for a given output (in other words, what is the minimal content that is needed to conduct a simulation with a given out in mind).
%
%\item Easiness to write format translators
%
%\item Input-output sequence in multiscale modeling (minimize information loss while maximizing the elimination of parameters)
%
%\item Standardize ways to define and establish minimal input deck quality in mesoscale applications
%
%\item Robust and flexible parsing of input data on demand (perhaps like in the Ovito atomic visualizer)
%\end{itemize}

\subsubsection{How do different \mapps fit within the Integrated Computational Materials Engineering (ICME) effort?}
\label{sec:icme}

\medskip
The notion of a family of \mapps spanning the mesoscale falls within the scope of the ICME paradigm \citep{Horstemeyer2018-lh}. The development of \mapps should therefore be done in a manner that acknowledges efforts made within ICME communities (coordination, best practices, etc.), and identify gaps in ICME that could be filled by \meso (or vice versa).

Also the inputs and outputs of \mapps should be consistent with successful computational tools within the materials modeling domain, such as CALPHAD (CALculation of PHAse Diagrams) \citep{lukas2007}, and other software tools focused on mechanical deformation (e.g., the UMAT protocol in the Abaqus finite element code).

%\subsubsection{Challenges}
%\begin{itemize}
%\item \mapps are generally not readily available for on-demand use in community codes thereby limiting their usage. Further, the lack of centralized curation of \mapps subject to provenance control makes reproducibility of others' work difficult or impossible.
%
%\item Often research codes are designed to perform well for particular classes of research problems, with limited transferability to other problems. The range of applicability of codes is often not widely known or documented.
%
%\item Repository of existing data/tools that already exist for materials research
%
%\item Research codes are not normally developed using state-of-the-art software engineering practices (DevOps), including version control, implementation verification, continuous integration and deployment, scalability, and adaptability (e.g., ease of incorporating new features).
%\end{itemize}

\subsection{Experimental Validation}
\label{sec:theme2:experiment}

\subsubsection{To what extent and in what forums are there interactions between experimentalists and modelers of mesoscale phenomena?}

\medskip
As in all collaborations, interactions are driven in part by financial support (federal and industry funding, nantional lab facilities, etc.), but also through personal acquaintance. When researchers do not know each other it can be difficult to obtain specific experimental information and data and/or access to computational tools, write collaborative proposals, etc.

To facilitate access to experimental and/or simulation data it is recommended to follow the August 25, 2022 memo by the Office of Science and Technology Policy (OSTP) in the Executive Office of the President of the United States tasking federal funding agencies to ``develop strategies to make federally funded publications, data, and other such research outputs and their metadata findable, accessible, interoperable, and re-useable, to the American public and the scientific community in an equitable and secure manner'' \citep{ostpmemo22}. Public access policies should be transparent, open, secure, and expeditious. In response to this call, a grassroots standard for metadata accompanying computational materials science datasets, called the \textit{Materials Core Metadata} (MatCore) standard is currently under development \citep{greenberg2025}.
%Metadata should include at minimum author names, affiliations, sources of funding, digital persistent identifiers (researchers should obtain per NSPM-33 Implementation Guidance); date of publication; and a unique digital persistent identifier for the research output

A secondary challenge is making it possible for modelers and experimentalists to understand each other to enable for integrated co-design of experiments, i.e., the design of experiments that can be used to parameterize and/or validate \mapps. There a variety of ways to achieve this, such as
\begin{itemize}
\item Training workshops in which modelers and experimentalists work together on the same target/physical problem. PI/senior personnel commitment is important.
\item In collaborative projects, modeling students should be involved in experiments, and experimentalists should be involved in simulations.
\item For experimental co-design, experimentalists need to know the limitations of models, and modelers need to know what can be measured in experiments.
\end{itemize}

\subsubsection{What overlap exists between measurements in state-of-the-art mesoscale experiments and \mapp predictions? What information and measures are required from experiments for \mapp parameterization and validation?}

\medskip
With recent advances in experiments (see Section~\ref{sec:background}) and increasing computational power, substantial overlap exists between measurement and computation. However, work remains to capitalize on this by extracting information from data; models tend to do forward prediction  whereas experiments provide results that require inverse models to make contact. This requires an agreement on measures to compare on (displacements versus strains, lattice rotations versus dislocation densities, etc.).
%\begin{itemize}
%\item Work to be done on extracting information from data; models tend to do forward prediction whereas experiments give you results $\Rightarrow$ couple them to do inverse models on experiments to extract more data.
%\item This requires strong trust in forward predictions of models, which becomes a problem of validation (breadth + depth)
%\item Need agreement on fundamental ground to compare on (displacements vs. strains; lattice rotations vs. dislocation densities)
%\end{itemize}

%\et{Don't understand the following:}
% \begin{itemize}
% \item Need to Work on Multiscale Aspect of Models to Reduce Dependencies on Costly Experiments for Every Step
% \begin{itemize}
% \item Don’t yet have validation steps at every step (e.g., MD simulations at $10^7$ and above)
% \item Better integration of modeling into the ‘rough cuts’ of experimental design
% \item Run first order simulations to see where things are; leverage existing modeling frameworks
% \end{itemize}
% \end{itemize}

Some models are post-dictive (i.e., they describe experimental observations) not predictive. To move past this, a dual-pronged approach is necessary. First, high-fidelity multiscale models should be developed with due cognizance to coarse-graining key unit processes (e.g., dislocation mechanics, twinning, phase transformation) from sub-scale models. The results from coarse-grained models should be amenable to rigorous, validation of well-defined metrics against finer-scale results. Second, double blind studies, such as the Sandia Fracture Challenge \cite{Boyce2014-gh}, are important. Such problems should cover a sufficient range of complexity and yet be clean enough to simulate). Identification of such problems for validating \mapps is of great interest. Further {\it Gold Standard} validation datasets for blind studies need to be checked for repeatability and reproducibility (do different operators, techniques, etc., capture the same key phenomena).

%\et{Don't understand the following:}
%\sj{I commented the bulleted items}
% \begin{itemize}
% \item Validation versus Extensibility
% \begin{itemize}
% \item Modelers really need local values (e.g., strain fields) – stress-strain at continuum scale is a usual metric, drill down to texture/its evolution, evolving data like twin formation, dislocation structure, etc.
% \item Resolution matters in validation  --- temporal, spatial
% \item Intermediate data / e.g., data evolving in time
% \end{itemize}
% \end{itemize}

To the extent possible, experiments should provide fundamental measures rather than derived quantities (e.g., displacements as opposed to strains). When this is not possible, it is important to clearly transfer the post-processing process and assumptions. The key is for modelers and experimentalists to closely work together on understanding each others' ``language.'' Data analysis should be standardized to ensure meaningful comparison.

Based on the above aspects, the participants made the following recommendations for design of experiments for \mapp validation:
\begin{itemize}
%\item Paradigm shift with increased computational power to very large mineable datasets that can serve different modelers
\item Save as much data as possible: there may be unknown unknowns controlling the observed phenomena.
\item Use well-defined geometries and clear boundary conditions.
\item Characterize the initial stage of the material with as much detail as possible (this is a particular challenge for experimentalists).
\item Provide 3D datasets with high spatio-temporal resolution if possible.
\item Perform robust UQ on the experimental data (and later on the modeling predictions).
\end{itemize}

\subsubsection{Can a canonical set of standardized problems be defined that can be studied experimentally and serve as validation benchmarks for mesoscale modeling approaches? How can this be done objectively? How does this vary across \mapp classes?}

\medskip
At the heart of an infrastructure aimed at the development and validation of predictive \mapps is a set of carefully selected and designed experiments. These experiments constitute a \emph{canonical} set of problems to test \mapps predictions. The selection of canonical experiments must be done with great care as the generation of high-quality experimental datasets requires a great deal of time and resources. Simple problems are valuable as they are well-defined as easier to model, but provide only a limited test of the scope of \mapp functionality. For example, a tensile test oriented for single slip provides information on hardening on one family of slip planes, but not on interactions between slip planes. A recommended approach is to design canonical problems of systematically increasing complexity, e.g., single crystal single species $\to$ single crystal multispecies $\to$ bicrystal $\to$ alloying effects, etc. 

Regardless of the experimental test, it is important to collect sufficient statistics in order to enable robust UQ. This can be accomplished by performing high-throughput ``rough cut'' experiments to gather statistics followed by expensive, highly accurate experiments to obtain high resolution spatio-temporal fields data to probe \mapp predictions.

Canonical experiments should be selected based on academic, research and commercial interests in order to motivate buy-in with an aim for infrastructure sustainability (as discussed in Section~\ref{sec:sustainability}). This includes consideration for the complete chain from processing to microstructure to material properties. In particular, given the current increasing interest in high-entropy alloys (HEAs) , additive manufacturing techniques, and materials under extreme conditions (high temperature, high strain-rate, hypersonic fluid--structure interactions, etc.), these areas should be prioritized in order to meet the demand for predictive \mapps and generate interest in supporting a \meso validation infrastructure.

%\begin{itemize}
%\item Canonical Problems Should:
%\begin{itemize}
%\item Test Robustness \& Extensibility
%\begin{itemize}
%\item Datasets are \$\$\$\$ --- where do we spend our time and effort?
%\item Simple systems are valuable; necessary but not sufficient
%\item Datasets gradually building up system --– data single crystal, bicrystal, systematic addition of alloys, etc.
%\item Transferability from a simple system with relevant physics to a more complex system
%\item Need the statistics from the experiment – provide the uncertainties
%\item Robust test statistics and high spatio-temporal resolution are inversely related
%\item Can use `rough cut' high throughput tests for statistics + modeling to guide where to high \$\$ spatio-temporal resolution experiments (will also probe robustness)
%\end{itemize}
%\item Motivate and Keep Infrastructure Sustainable
%\begin{itemize}
%\item Motivate `buy-in' by demonstrating value quickly $\Rightarrow$ Materials Design
%\item Pick demonstration material(s) that build up re above, eye towards broader frameworks/approaches
%\item Processing-structure-properties framework
%\item E.g., Additive manufacturing, Lightweighting
%\item E.g., extreme conditions: High temperature, high strain rate, hypersonics / etc.
%\item Framework development (software capabilities, etc.)
%\end{itemize}
%\end{itemize}
%\end{itemize}

\subsubsection{What experimental data processing and analysis tools exist, and which of these would it be useful to share across the community? How specific are the tools to particular experimental methods? How would it be best to curate and share such tools?
}

\medskip
Major advances in high-precision experimentation probing material response at the mesoscale (see Section~\ref{sec:background}) open the prospect to targeted validation of \mapp capabilities. Beyond faacility-scale beam line resources, this include techniques such as BCDI, high-resolution DIC, EBSD, lab-scale HEDM \et{Ashley citation?}, and nano 3D tomography. 

The data obtained from such methods must be processed into a form that can be used to validate \mapp predictions. Examples include the construction of displacement and strain fields from DIC measurements, and extraction dislocation slip from EBSD maps. Currently experimental postprocessing codes are largely being developed within in individual experimental groups and are not widely shared. This results in a lack of standardization in output formats, and constitutes a missed opportunity for collaborative community effort to develop and verify the correct implementation of such tools. A \meso infrastructure is an ideal setting for curating and sharing experimental postprocessing codes to enable collaboration among experimentalists with input from modelers on their needs.

%\begin{itemize}
%\item Microstructural and Mechanics Data Analysis Tools
%\begin{itemize}
%\item Open source DIC codes (e.g., local codes, analysis of resultant displacement/strain fields), EBSD analysis, etc.
%\end{itemize}
%\item Quantifying Information from Reconstructions
%\begin{itemize}
%\item E.g., Slip from an EBSD map –-- how do you turn this into a plot to compare models against?
%\end{itemize}
%\item How to Best Curate and Share?
%\begin{itemize}
%\item Several tools are privately constructed; useful to share and compare
%\item \textbf{Publicly available tutorials are needed}
%\item Avoid putting all eggs in one basket
%\item How do you know when a tool is wrong without comparing?
%\item What happens if tool gets acquired and is no longer open source?
%\item \textbf{Provenance Builds Trust}
%\item Original collection, changes, etc.
%\end{itemize}
%\item Sharing Data and Codes – How to Handle the Work/Questions that come from that?
%\begin{itemize}
%\item Collaborate with modeling people and target variables they really need from experiment
%\item For codes, move personal communication from email/text to public message boards (Mtex, github message boards) to create public spaces for troubleshooting communications
%\end{itemize}
%\end{itemize}

\subsection{Cyberinfrastructure vision and design}
\label{sec:theme3:collaboration}

\subsubsection{Will \meso include a code repository?}

\medskip
\meso will set up a GitHub organization to provide a secure and collaborative space for a trusted set of developers to share their work, and for general users to access the latest codebase and features. In addition, Galaxy specific tools will be deposited in the Galaxy Tool Shed \cite{re:g-toolshed} which is publicly accessible, and hence will be incorporated as part of Galaxy distributions. Galaxy does not provide at present revision control, so only latest versions will be made available. The GitHub organization will provide version control. In order to encourage users to use both resources effectively, \meso will provide Git technical support services for non-experts and training (see Section~\ref{sec:training}).

%\begin{itemize}
%\item Galaxy includes a ``tool shed'' where codes are stored. It is possible to store different versions of the same code, but Galaxy does not provide revision control.

%\item For long-term support of codes, \meso will include revision control through a GitHub organization.

%\item To encourage users to use this resource, \meso will provide Git technical support services for non-experts and training (see training below).
%\end{itemize}

\subsubsection{What standards and translation tools will be needed to connect \mapps with experimental outputs with a workflow manager such as Galaxy?}

\medskip
Translation tools between the many \mapps applications, their formats, and experimental data sets and data generation facilities will be provided within Galaxy. Before ingestion into the workflow manager, data needs to be typed according to formats that are pre-defined and recognized in Galaxy. As part of the workflow, translation tools are provided. This method provides for an easily extensible set of formats, as well as protocols for creating new data types and making them compatible with other tools. There is a great deal of experience in the Galaxy and the bioinformatics community in this regard, and we will follow the same development methods here. \meso personnel will work with researchers to determine whether new data types are necessary and provide support in developing the set of translation tools necessary to integrate the new data into existing workflows. By separating the code and data from workflow integration, it is possible to build flexibility into the system, which would be difficult to require otherwise from domain scientists. However, it will remain necessary to maintain a continuous unit/model testing framework for any new or modified \mapp (or new and modified experimental reduction tools), including translation tools, to pass through a verification layer to ensure integration is achieved. Again, separating the code and data layers from the translation layer will simplify the verification process.

%\begin{itemize}
%\item In Galaxy, ``type'' refers to the type of file, like atomic configuration. Each type can be stored in different formats, like XYZ, LAMMPS dump file, etc.

%\item It would be a benefit to define some standard formats, using existing formats whenever possible. If users choose to use these, new translators would not be needed. Otherwise either the user or a \meso staff person would need to write a translator. Not having to write a translator is a motivation to adopt existing standards.

%\item Continuous unit/model testing framework --- any new/modified \mapp, new/modified experimental reduction tool, or new/modified translator will pass through a verification layer to make sure that it works with the platform.
%\end{itemize}

\subsubsection{What resources will be needed to enable the community to participate actively in contributing to \meso and participating in validation challenges?}

\medskip
Attention will be paid to developing resources that will allow the community to actively participate in contributing to \meso, and in validation challenges. The project is expected to provide a set of positive incentives to promote code/data contributions. They will include the requirement that in order to use \meso computational resources or receive dedicated support, users will need to upload publicly available content (usage will remain free on users’ own resources). Direct funding for code development that will be uploaded to \meso is envisaged. For example, funding for graduate students or postdocs, paid internships at participating institutions and industry partners for graduate students/postdocs of \meso participants for work related to \meso content, or even funding for code contributors to enable them to continue to support their already contributed codes. In addition, the possibility of dedicated technical support (software engineering/data science) and domain specific support (experts in mesoscale modeling and experiments) from \meso staff may induce outside researchers to participate and contribute.

Other, less traditional methods will be explored. A ``Spotify''-type revenue sharing model (either in funds or in-kind support) will be considered for developers based on \meso content used by commercial users of the platform. Funds from corporate users may be pooled to support developers, as is currently done to support the development of open source software. 

Indirect incentives may also be adopted to promote contributions. For example, journals and funding agencies may view contributions to \meso favorably at submission time, motivated by contributions to an already existing infrastructure. The project will provide reference documentation for codes, data, and shared workflow histories that can be publicly discoverable and cited in publications.

\subsubsection{What hardware resources will be required to support the computational and data storage requirements of the project?}

\medskip
\meso will have dedicated hardware resources for its users to host codes, workflow manager and associated products, and curated experimental datasets. It will also feature visualization tools, and the capability to manage VR communications (see Section~\ref{sec:vr}). Community allocation at NSF supercomputing installations will be requested as well. Galaxy can automatically decide on what hosts (local or remote) to run portions of the workflows depending on the computational resources that any given task is required. Such a communication and offloading of tasks to NSF nodes is already routine in some place, e.g., at the Minnesota Supercomputing Institute \et{Jorge: Was the intent here UMN or MSI?} that can be leveraged to establish the connections and the allocations on behalf of the community. Hardware will include a mix of traditional CPU resources, GPU accelerators, and the switching infrastructure to manage parallel computation. Parallel processing will be enabled with InfiniBand connectivity across nodes. Parallel storage needs to be supplied. It is anticipated that only curated data sets of benchmark nature will be stored at \meso resources. Other raw data will be stored at the facilities in which it is being produced. Galaxy pulsar servers can be set up at those facilities so that the raw data is discoverable and accessible to user workflows. In the originating facility does have some computational resources, workflow tasks such as data reduction and analysis can be done on site. Galaxy central would only received the reduced data, with pointers to the raw data maintained.

%\begin{itemize}
%\item \$3M = CPU Petaflop/ \$0.5M GPU Petaflop + storage
%\item HPC graphics piped to users (run Paraview on \meso resources)
%\item Question: Is it possible to host at existing NSF centers or buy dedicated resources there?
%\end{itemize}

\subsubsection{What forms of VR and augmented reality have the potential to contribute to researcher collaboration, productivity, and education?}
\label{sec:vr}

\medskip
VR will be a part of the infrastructure as augmented reality has the potential to contribute to researcher collaboration, productivity, and education (VR meetings and VR based educational content). More advanced uses include virtual experimental execution at remote facilities, with resulting data flows integrated into the workflow manager. Other novel research modalities that can be enabled by VR resources include computation or analysis simultaneously with visual representation, and visual based computation (launching tasks and controlling workflows through a visual interface). Both research and educational VR services will be publicised, and leveraged for educational purposes. 

%\begin{itemize}
%\item VR meetings
%\item VR-based educational content
%\begin{itemize}
%\item VR lab where someone can learn how to run an experiment, e.g. virtual TEM lab, virtual X-ray diffraction
%\item Visalization
%\end{itemize}
%\item Computation with visual input (e.g. visualize a grain structure, rotate a grain and have the rest recomputed)
%\item VR for 3D workflow management
%\item Service to generate VR content based on \meso participant research to publicize \meso activities and for education purposes.
%\item Question: Can VR content be translated to other 3D visualization frameworks, such as a planetarium?
%\end{itemize}

\subsubsection{What are workforce training needs for advanced mesoscale modeling and experiments and the connection between them, such as data science, UQ, machine learning, and software carpentry?}
\label{sec:training}

\medskip
The research infrastructure contemplated will contribute to work force training in an advanced environment. The skills that are necessary to operate and utilize \meso include classical software engineering expertise (software engineering, software carpentry, development and systems operation), all in a HPC environment with resources made available nationwide. In addition, fast emerging modalities of data based computation are quickly being adopted by the materials science community, including UQ tools, and machine learning.

\meso will provide training to develop these skills through the production of  education and training material, delivered online through YouTube videos and web tutorials. Exemplars from code contributed to \meso will lead to tutorials. In addition, Coursera like, or MasterClass sets of courses taught by domain experts (on an on demand framework) will be provided in conjuction with the capabilities (code, data, and workflow histories) available on the platform. This material can then be integrated into curriculum development efforts at the participating institutions. Summer validation schools will be held, mostly geared towards students, as well as short courses or tutorials at related conferences. A forum at matsci.org will be created as part of this training effort.

Finally, we will develop an AI support bot by fine tuning existing LLMs on verbal and text input related to the tools available on the project. Ideally the bot will keep abreast of new papers, journal abstracts, preprint server information that is generated and it relates to the materials community around \meso.

%\begin{itemize}
%\item Needed skills:
%\begin{itemize}
%\item DevOps/software carpentry/software engineering
%\item HPC
%\item Data Science
%\item Machine learning
%\item UQ
%\end{itemize}
%\item Methods for providing these skills:
%\begin{itemize}
%\item Education and training material (YouTube videos, web tutorials)
%\begin{itemize}
%\item Develop exemplars from code contributed to \meso as tutorials
%\item Coursera-like or MasterClass set of courses taught by domain experts (on-demand framework) – need to be careful not to compete with education projects like nanoHub.
%\item Curriculum development to integrate within existent courses
%\end{itemize}
%\item Summer validation schools (teach skills)
%\item Short course/tutorial at conferences
%\item Technical support in the skills areas
%\item Forum on matsci.org
%\end{itemize}
%\item AI support bot
%\begin{itemize}
%\item Fine tuning GPT3 on all verbal and text input to the project, and other domain relevant content (conferences, youtube, arXiv, journal abstracts, %\dots%)
%\end{itemize}
%\end{itemize}

\subsubsection{What are the implications of FAIR principles in ensuring easy and equitable access to codes and data?}

Any cyberinfrastructure for community use must adhere to FAIR (Findable, Accessible, Interoperable, and Reusable) principles \citep{Wilkinson2016-es, Scheffler2022-wp, Walsh2024-mw} to ensure easy and equitable access to code and data. For the purposes of \meso we distinguish three components: data (mostly of experimental origin), codes (data reduction, analysis, modeling and simulation, visualization), and workflow manager infrastructure. 

Since the workflow manager is based on Galaxy, we will rely on the extensive set of tools for installation, curation, dissemination and training that Galaxy already has. Any tool which is deployed as part of \meso will be deposited in the Galaxy ToolShed \cite{re:g-toolshed} following their existing protocols. The tools then become publicly available with any installation of Galaxy from the public domain. Installations are currently managed through the automation engine Ansible \cite{re:ansible}

The workflows developed by users can be shared directly, or in the form of histories of already run workflows. They are not only resusable, but can be used to exaclty reproduce prior calculations. All data, tools, and metadata are kept as part of the history, and hence calculations can be reproduced without any additional work. Histories can be shared, or published in order to determine the degree of accessibility by users of the platform or the public in general. Galaxy also provides a mechanism for shared data libraries, and connectors to other established repositories on the Web.

Interoperability is guaranteed through the initial definition of data types, the typing of data and code as it is being produced, and its integration into the workflow manager. The set of data types is extensible and can be updated as new types of data or modalities of research emerge. However, users or \meso staff will be responsible for writing any connection tools that are required for the data types to be part of the workflows, and hence interoperable.

Data can be categorized in two types: raw data that may be too large to efficiently move, and derived data that will be held on Galaxy servers or at other supercomputing installations. Concerning the former, Galaxy pulsar servers will be deployed at data producing facilities so that the data is accessible to workflows (the data is indexed with Galaxy, and the reduction or analysis code is moved to the facility as part of the workflow, without transporting the data). This process is transparent to uses. Reduced or derived data, if smaller in size, can be move to either Galaxy servers or to HPC installations where models will be run. Galaxy can also be configured so that large workloads are executed remotely. 
%Centers in NSF/ACCESS grant community allocations that are used to execute workflow manager initiated computations \et{Jorge: Sentence unclear}. 
Any data traffic back and forth is also handled by the workflow manager, and the process is transparent to the user.

The third component (codes) will be uploaded to the Galaxy toolshed if in production, while various prior versions, or current development will be available in the GitHub organization set up for \meso. All \meso developed content will be open source, and all users will be encouraged to contribute content as open source. License information for all content will be explicitly provided (see \cite{re:spdx}).

Key code components or derived data will be made more broadly discoverable through metadata standardization \cite{re:matcore,re:prism}. 

%\begin{itemize}
%\item Findable: Metadata (look into existing metadata standards, such as work from PRISMS and MatCore)
%\item Accessible: Licensing
%\begin{itemize}
%\item Provide license information for all content (SPDX)
%\item All \meso developed content will be open source
%\item Encourage contributed content to be open source
%\end{itemize}
%\item Interoperablity: standard formats
%\item Reusability: Persistent IDs
%\end{itemize}

\subsubsection{What are possible approaches for the long-term sustainability of the effort, such as the formation of an industry consortium?}
\label{sec:sustainability}

\medskip
Long-term sustainability of community platforms is a long standing issue, not just in materials science. Public funding serves to initiate a number of activities, which are then meant or asked to become self sustaining with user base support. However, as long as the user base remains mostly academic, with sources of funding narrowly tailored to specific research projects, there are just no viable mechanisms for support of broad, multi-institution environments. Examples of successful, long duration projects, all include direct and ongoing government funding. This is the case both for international laboratories (e.g., CERN, Antartica Research Station, astronomical observation telescopes in orbit or on the ground, even the International Space Station), or the network of US National Laboratories. Closer to materials science, we mention the large National Laboratories that provide fee based services (e.g., Argonne National Laboratory, the National High Magnetic Field Laboratory, and many others), or national supercomputing installations. In the former case, user fees provide a small fraction of the operating budget of the Laboratory. In the latter, usage is largely free, with deployment and operation costs born entirely by the funding agencies.

Even closer to this effort, we note that development of the software platform Galaxy, created more than two decades ago, is sill largely supported by US, European, and Australian public funding sources \cite{re:g-funding} . Similarly, NanoHub which was launched in 2002 relies largely on public support for its operation, with a last round of major funding in 2019 (US NSF) \cite{re:nanohub-funding}. The OpenKIM project \cite{tes11} also continues to rely on government funding.

In addition to the proposed \meso (see Section~\ref{sec:aims}) as a founding event, and standard activities such as revenue from industrial affiliates,  or the provision of services (consulting, training workshops and materials), will be explored. Staff in \meso will be made available to participate, part time, as funded personnel on outside projects. Their expertise and vetted credentials will greatly accelerate other funded research that would utilize the infrastructure, or overlap with the methodology maintained by the center. Their services can also be utilized for knowledge transfer, and postdoc or graduate student training in affiliated institutions. Such a participation would also indirectly benefit  \meso, as information and knowhow from other research projects are made part of the common knowledge base, and the project remains integrated with other research activities nation wide. Each staff member could participate in two or three projects at any given time. Some institutions offer overhead arrangements to a center of this type, so that the combination would go a long way towards making the personnel part of the center self-sustaining.

Other components: experimental and supercomputing facilities that will enable the sharing of information and workflows are already funded independently, and it is assumed here that they will continue to be. Community allocations of computational resources at HPC installations will be managed by \meso, and they are, at the moment, free. Periodic proposals for computational hardware and operation will be necessary, and perhaps some contribution to the complement of staff. If, however, there is sufficient demand for staff integration into other outside projects, it is anticipated that funding agencies would see the need to provide some common base of support to ensure the continuity of \meso.

%\begin{itemize}
%\item Classy advertising 
%\item \meso Manager (or another designated person) will lead and coordinate sustainability efforts. 
%\item Revenue from industrial affiliates:
%\begin{itemize}
%\item Work for hire services (consulting, simulations, …)
%\item Training/courses for industry professionals (like IPRIME)
%\item Potential affiliates: GE, car industry, aerospace companies, additive manufacturing startups, mechanical experimentation, defense industry, software companies
%\end{itemize}
%\end{itemize}

\section{Discussion and Conclusions}
\label{sec:discussion}

This paper describes a proposed research infrastructure \meso aimed at advancing the predictive capability of \mapps (mesoscale applications) for structural materials through a systematic co-design of validation experiments and rigorous UQ on both the experimental and simulation sides. A broad strokes introduction is presented in Section~\ref{sec:aims} followed by detailed best-practice recommendations emerging from an NSF workshop dedicated to this topic described in Section~\ref{sec:community}. This effort can serve as a prototype for many other scientific domains in which models calibrated against experiments are used to predict physical phenomena.

\et{Need to expand}

\section*{Acknowledgements}
This work was supported in part by the National Science Foundation under Grant No.\ 2231655. The authors thank the workshop participants for their contributions and helpful discussions.

%% The Appendices part is started with the command \appendix;
%% appendix sections are then done as normal sections
%\appendix

%\section{Sample Appendix Section}
%\label{sec:sample:appendix}
%Lorem ipsum dolor sit amet, consectetur adipiscing elit, sed do eiusmod tempor section \ref{sec:sample1} incididunt ut labore et dolore magna aliqua. Ut enim ad minim veniam, quis nostrud exercitation ullamco laboris nisi ut aliquip ex ea commodo consequat. Duis aute irure dolor in reprehenderit in voluptate velit esse cillum dolore eu fugiat nulla pariatur. Excepteur sint occaecat cupidatat non proident, sunt in culpa qui officia deserunt mollit anim id est laborum.

\appendix
\setcounter{table}{0}
\renewcommand{\thetable}{A\arabic{table}}

\section{Science Gateway infrastructure and management}
\label{sec:sgmanagement}

\et{Consider removing, shortening or integrating this appendix with the main text. I already commented out some of it.}

A Science Gateway is a well understood framework that integrates data, software components, and computational resources in an intuitive, browser-based work area, and that supports visual/point and click management of workflows as required and specified by the community. A gateway will provide a single point of access for the software and hardware resources required for processing, analysis, and simulation of acquired data, as well as integrate with public benchmark data sets for further research and instruction. It will provide a mechanism for rapid development and testing of new tools as they become available, for their deployment globally to domain users, and for supporting users of new codes by facilitating interactions between users and the code developers. A gateway will serve three distinct functions: (1) enabling users at advanced facilities who are collecting and processing data to use state of the art reconstruction and analysis tools, (2) providing a platform that simplifies creating, testing and improving contributed analysis and simulation codes, and (3) enabling remote users access to data sets and the best available codes, including modeling and simulation. For the discussion that follows, we will assume that the Science Gateway is built around the workflow manager Galaxy  \cite{re:goecks10,re:vinals21}. Other workflow managers exist at present, but Galaxy has proven capabilities, and large and active user and developer communities. Therefore, it is simpler to refer to capabilities needed for the present effort back to Galaxy's already existing features.

Management of a Science Gateway requires a dedicated set of professionals. This would include information technology (IT) personnel to work collaboratively with IT groups in the various nodes, domain specific research scientists that would know the needs of the community of users, and workflow management experts. Oversight would ideally rest in a committee drawn from one or a few of the external user groups already existing at the data producing facilities. This would allow a natural avenue for communication with the community, and for the gateway to be responsive to its needs. Community HPC allocations can be easily requested to support the modeling and simulation needs of the gateway, which can also be managed by the oversight group.

Data and history archival naturally takes place within the gateway, as well as public access to both. However, software development related to gateway development needs to be addressed separately. The workflow manager Galaxy, for example, allows archival of all its tools in the Galaxy Toolshed \cite{re:g-toolshed}, which is publicly available. In addition, GitHub Organizations provide another avenue for gateway code dissemination, including training materials. We mention that the Galaxy community already maintains a very large collection of online training materials \cite{re:g-training}, as well as coordinating a wide variety of dissemination events, including at topical conferences. This is largely a community driven effort, and a good model to emulate (or eventually join) by the Science Gateway for Materials Research.

\section{Workshop Agenda and List of \meso workshop participants}
\label{app:participants}
%\et{COMPLETE}
\paragraph*{\bf Agenda:} The \meso workshop comprised invited talks and workgroup meetings focused on three themes:
\begin{itemize}
\item {\bf Theme 1:} Nature of mesoscale models
\item {\bf Theme 2:} Experimental validation
\item {\bf Theme 3:} Cyberinfrastructure vision and design
\end{itemize} 

It included invited participants identified as leaders in mesoscale modeling, experimentation, and cyberinfrastructures, cf.~Table \ref{tab:themepanels}. 

\begin{table}[h!]
    \centering
    \begin{tabular}{|c|c|c|}
        \hline
        Theme 1 & Theme 2 & Theme 3 \\ \hline
        {\bf Jim Warren}       &   {\bf Jaafar El-Awady}      &     {\bf Ellad Tadmor}    \\ \hline
        {\bf Jaime Marian}       &  {\bf Sam Daly}       &  {\bf Jorge Vin{a}ls}       \\ \hline
         Miguel Bessa      &   Suraj Ravindran      &   Krishna Garikipati      \\ \hline
         Tim Truster      &    Eric Hintsala     &      Mingjian Wen   \\ \hline
        Brandon Runnels     &  Marko Knezevic       &    Giacomo Po     \\ \hline
         Ill Ryu      &     Leslie Mushongera    &     Shailendra Joshi    \\ \hline
         Shuozhi Xu      &   Theocharis Baxevanis      &  Youhai Wen       \\ \hline
          Saurabh Puri (online)     &   Darren Pagan (online)      &    Surya Kalidindi     \\ \hline
          Nikhil Admal (online)     &    Justin Wilkerson (online)     &  Somnath Ghosh (online)       \\ \hline
               &         &   Graham Candler (online)      \\ \hline
    \end{tabular}
    \caption{Thematic discussion groups. Leads are highlighted in bold.}
    \label{tab:themepanels}
\end{table}

The detailed agenda for the two-day workshop is given in Tables \ref{tab:day1} and \ref{tab:day2}
\pagebreak

\begin{table}[h!]
\centering
\begin{tabular}{|c|c|p{8.2cm}|}
    \hline
    \textbf{Start Time} & \textbf{End Time} & \textbf{Agenda} \\ \hline
    7:00 a.m. & 8:00 a.m. & Breakfast \\ \hline
    8:00 a.m. & 8:20 a.m. & Welcome \& Opening Remarks \\ \hline
    8:20 a.m. & 8:50 a.m. & \textbf{Keynote: Prof. Graham Candler} \\
               &          & Code Validation Studies for Hypersonic Aerodynamics \\ \hline
    8:50 a.m. & 9:15 a.m. & \textbf{KnoMME Infrastructure} - Ellad Tadmor \\ \hline
    9:15 a.m. & 9:35 a.m. & \textbf{Theme 1:} Dr. Jim Warren \\
               &          & Mesoscale Modeling is Hard \\ \hline
    9:35 a.m. & 9:55 a.m. & \textbf{Theme 1:} Prof. Jaime Marian \\
               &          & Status and prospects of mesoscale materials modeling in guiding materials design and manufacturing \\ \hline
    9:55 a.m. & 10:15 a.m. & Discussion / Q\&A \\ \hline
    10:10 a.m. & 10:30 a.m. & Break \\ \hline
    10:30 a.m. & 10:50 a.m. & \textbf{Theme 2:} Prof. Surya Kalidindi \\
               &          & Integration of Experimental and Simulation Datasets Using AI/ML \\ \hline
    10:50 a.m. & 11:10 a.m. & \textbf{Theme 2:} Prof. Ashley Bucsek \\
               &          & Synchrotron experiments - State of the art, future capabilities, and ongoing challenges \\ \hline
    11:10 a.m. & 11:40 a.m. & Discussion / Q\&A \\ \hline
    11:45 a.m. & 12:45 p.m. & Lunch \\ \hline
    12:45 p.m. & 1:05 p.m. & \textbf{Theme 3:} Prof. Jorge Vinals \\
               &          & Galaxy: A workflow manager to link data sources, simulation, and analysis in a delocalized environment \\ \hline
    1:05 p.m. & 1:25 p.m. & \textbf{Theme 3:} Prof. Ellad B. Tadmor \\
               &          & Lessons learned from the OpenKIM cyberinfrastructure \\ \hline
    1:25 p.m. & 1:55 p.m. & Discussion / Q\&A \\ \hline
    1:55 p.m. & 2:15 p.m. & Break \\ \hline
    2:15 p.m. & 5:15 p.m. & Panel Breakout Sessions (Themes 1-3) \\ \hline
    5:15 p.m. & 6:00 p.m. & Extra Time - Theme leaders may use this time to summarize outcomes and prepare for the second day \\ \hline
    6:30 p.m. & 8:30 p.m. & Dinner \\ \hline
\end{tabular}
    \caption{Day 1 (12 January 2023) Agenda}
    \label{tab:day1}
\end{table}

\pagebreak
\begin{table}[h!]
    \centering
\begin{tabular}{|c|c|p{8.2cm}|}
    \hline
    \textbf{Start Time} & \textbf{End Time} & \textbf{Agenda} \\ \hline
    7:00 a.m. & 8:00 a.m. & Breakfast \\ \hline
    8:00 a.m. & 8:15 a.m. & Opening Remarks \\ \hline
    8:15 a.m. & 9:15 a.m. & \textbf{Theme 1:} Panel Presentation \newline (Jim Warren, Jaime Marian) \\ \hline
    9:15 a.m. & 10:15 a.m. & \textbf{Theme 2:} Panel Presentation \newline (Jaafar El-Awady, Sam Daly) \\ \hline
    10:15 a.m. & 10:30 a.m. & Break \\ \hline
    10:30 a.m. & 11:30 a.m. & \textbf{Theme 3:} Panel Presentation \newline (Jorge Viñals, Ellad B. Tadmor) \\ \hline
    11:30 a.m. & 12:00 noon & Workshop Co-chair Summary \\ \hline
    12:00 noon & 1:00 p.m. & Adjournment. Lunch - to go \\ \hline
\end{tabular}
    \caption{Day 2 (13 January 2023) Agenda}
    \label{tab:day2}
\end{table}

%% If you have bibdatabase file and want bibtex to generate the
%% bibitems, please use
%%
 \bibliographystyle{elsarticle-num}
 \bibliography{hMESO}

%% else use the following coding to input the bibitems directly in the
%% TeX file.

% \begin{thebibliography}{00}

% %% \bibitem{label}
% %% Text of bibliographic item

% \bibitem{}

% \end{thebibliography}
\end{document}